\documentclass[10pt,conference]{IEEEtran}
\IEEEoverridecommandlockouts

\usepackage{geometry}
\usepackage{amsmath,amssymb,amsfonts}
\usepackage{algorithmic}
\usepackage{graphicx}
\usepackage{textcomp}
\usepackage{xcolor}
\usepackage{booktabs}
\usepackage{dsfont}
\usepackage{tikz}
\usepackage{amsthm}
\usepackage{adjustbox}
\usepackage{multirow}
\usepackage[linesnumbered,ruled]{algorithm2e}
\begin{document}

\title{
A Game-Theoretic Approach for PMU Deployment Against False Data Injection Attacks
}

%Detecting False Data Injection Attacks with One Additional PMU

\author{\IEEEauthorblockN{Sajjad Maleki\IEEEauthorrefmark{1}\IEEEauthorrefmark{2},
Subhash Lakshminarayana\IEEEauthorrefmark{1}, E. Veronica Belmega\IEEEauthorrefmark{3}\IEEEauthorrefmark{2}
Carsten Maple\IEEEauthorrefmark{4}}
\IEEEauthorblockA{
\IEEEauthorrefmark{1} School of Engineering, University of Warwick, Coventry, United Kingdom \\
\IEEEauthorrefmark{2}
ETIS UMR 8051, CY Cergy Paris Universit\'e, ENSEA, CNRS, F-95000, Cergy, France\\
\IEEEauthorrefmark{3}Univ. Gustave Eiffel, CNRS, LIGM, F-77454,  Marne-la-Vallée, France\\
\IEEEauthorrefmark{4}WMG, University of Warwick, Coventry, United Kingdom\\
Email: sajjad.maleki@warwick.ac.uk, subhash.lakshminarayana@warwick.ac.uk, \\veronica.belmega@esiee.fr, cm@warwick.ac.uk}
\thanks{This work has been supported in part by the PETRAS National Centre of Excellence for IoT Systems Cybersecurity through the U.K. EPSRC
under Grant EP/S035362/1 EUTOPIA and in part by PhD Cofund WALL-EE project between the University of Warwick, UK and CY Cergy Paris University, France.}
\vspace{-3em}} 

\maketitle

\begin{abstract}
Phasor Measurement Units (PMUs) are used in the measurement, control and protection of power grids. However, deploying PMUs at every bus in a power system is prohibitively expensive, necessitating partial PMU placement that can ensure system observability with minimal units. One consequence of this economic approach is increased system vulnerability to False Data Injection Attacks (FDIAs). This paper proposes a zero-sum game-based approach to strategically place an additional PMU (following the initial optimal PMU deployment that ensures full observability) to bolster robustness against FDIAs by introducing redundancy in attack-susceptible areas. To compute the Nash equilibrium (NE) solution, we leverage a reinforcement learning algorithm that mitigates the need for complete knowledge of the opponent's actions. The proposed PMU deployment algorithm increases the detection rate of FDIA by $36\%$ compared to benchmark algorithms.
\end{abstract}

\begin{IEEEkeywords}
Cybersecurity, False Data Injection Attacks (FDIA), Zero-sum Games, Nash Equilibrium, PMU.
\end{IEEEkeywords}

\section{Introduction} \label{Intro}
Phasor Measurement Units (PMUs) have become pivotal in power systems thanks to their increased accuracy, high sampling rate, and time-synchronised measurements. However, they must be carefully and sparingly deployed because of their high cost. As a result, optimal PMU placement that ensures full system observability has been widely investigated in the literature \cite{jamei2019phasor, elimam2021novel}.
% These PMUs are distributed throughout the power grid, forming a network of data sources. 

PMU measurements play a crucial role in the state estimation of power systems, providing essential data for their accurate operation. However, 
the communication technologies for PMU operations (communications channels and global positioning system for synchronization) introduce cyber security risks. In particular, the communication channel is susceptible to attacks aimed at manipulating PMU measurements through the injection of false data \cite{mousavian2014probabilistic}.

The presence of a PMU in a bus or its {adjacent buses} 
allows an operator to obtain the necessary measurements to determine the state of the bus. If all system states can be obtained through PMU measurements, then the system is considered fully observable.
Full observability of the system is necessary for state estimation of the power systems.  Consequently, many works including \cite{ghosh2017optimal, zhang2022attack} proposed different methods for optimal PMU placement to guarantee the observability of the system. In \cite{ghosh2017optimal}, 
the authors have introduced statistical criteria to determine the optimal locations for PMUs and employ a multi-criteria decision-making approach known as the analytical hierarchical process to optimize their placement.  In \cite{zhang2022attack}, a hierarchical process for enhancing power system observability is introduced, which proposes an iterative approach in which the operator adds a PMU at a time to guarantee the maximum observability of the system with the available number of PMUs. {In \cite{pei2020pmu} and \cite{yang2017optimal}, authors propose greedy algorithms for PMU deployment not only to achieve full observability but also to detect stealthy FDIAs on the Supervisory Control and Data Acquisition (SCADA) measurements. It's worth noting that these papers assume that the PMUs themselves are secure against FDIAs. Reference \cite{badrsimaei2023observable} utilizes a modified optimal PMU placement algorithm to develop a defense solution to attacks causing a surge in the electricity price due to the false data injected by adversaries.} {Authors of \cite{yang2017pmu} propose that PMUs themselves could be vulnerable to FDIAs. Consequently, they advocate for a new PMU placement method to mitigate such attacks. Their research demonstrates that increasing the number of PMUs reduces the probability of undetected FDIAs}. However, they do not consider the presence of a strategic attacker.

To the best of our knowledge, \cite{ghosh2023bi} is the first work to propose a game theoretic approach to tackle the optimal PMU placement problem in the presence of cyber threats {initiated by strategic attackers.} {Their methodology involves implementing a bi-level game in which the attacker's objective is to deviate the state estimation from the true value in a stealthy manner. On the other hand, the operator tries to optimally place the PMUs to maintain the full observability of the system and minimize the state estimation errors.} 
%{\color{red} [what are these costs?]} {\color{red} [game between who and who? what is full knowledge - of what? Why is this full knowledge hard to acquire? you have also not introduced what attack, or what security problem is considered.]}

Previous works have extensively investigated the observability of the system and {enhanced the robustness of PMU measurements against manipulation in state estimation by adversaries}. However, there remains a gap in the literature regarding research on mitigating the likelihood of successful FDIAs in the presence of a strategic attacker.{ Unlike \cite{ghosh2023bi}, which proposes a completely new PMU placement scheme to achieve robustness against FDIAs, this paper assumes that optimal PMU placement has already been made to ensure system observability. Our focus is on adding an additional PMU to the system to decrease the likelihood of successful FDIA through the resulting redundancy in measurements. Thus, our approach is more practical for securing power systems that are already starting to witness PMU deployment, mainly with the observability criterion in mind.
 
In this work, we assume that a strategic attacker aims to launch such an FDIA on a PMU's measurement that is not observable by any other PMU while having the biggest possible impact on the state estimation. We exploit a two-player zero-sum non-cooperative game \cite{lakshminarayana2021moving} to analyze the interaction between the attacker and defender and investigate the robust minimax defense solutions at the NE. The goal of the defender is to increase the FDIA detection rate against strategic attacks by adding redundant observing PMU to attack-prone measurements.  
Then, we propose a reinforcement learning algorithm drawn from multi-armed bandits, namely exponential weights for exploration and exploitation (EXP3), to show that the robust NE solution can be computed without the full knowledge of the game at the defender's end but via iterative interactions with the attacker. 

We analyze the impact of incorporating an additional PMU using a game-theoretic approach applied to the IEEE 14-bus system. This analysis involves comparing the outcomes with two benchmarks obtained by: (i) not adding any additional PMUs and (ii) randomly adding an extra PMU. A higher attack detection percentage of the proposed solution encourages the expansion of this work to larger systems which may require more than one PMU to reach a certain level of detection.

The key contributions of this paper are:
\begin{itemize}
    \item {Devising a minimax NE solution for obtaining a robust and low-cost defense strategy against strategic FDIA attacks by placing an additional PMU in the system.}
    \item Exploiting reinforcement learning to find the {NE solution} in an iterative manner without requiring any knowledge about the attacker's actions.
   \item We present extensive simulation results using the IEEE 14-bus system and compare the performance of the proposed algorithm against a benchmark technique. {Results exhibit an enhanced detection rate for strategic attacks while using our proposed method.}
\end{itemize}

The rest of the paper is organised as follows: Section \ref{Problem} describes the problem formulation; Section \ref{Solution} discusses the proposed solution while in Section \ref{result}, {our} numerical results are presented and discussed and, finally, Section \ref{conclusions} concludes the paper.
%%%%%
%%%%%
\section{Problem Formulation} \label{Problem}
In this section, we first introduce the power system model and optimal PMU placement problem, before delving into the issue of the increasing robustness of PMU measurements against FDIAs.
\subsection{Power system model} \label{sub:power sys}
Let the graph $\mathcal{E}=\{\mathcal{N},\mathcal{L}\}$ represent the power system, where $\mathcal{N}$ is the set of buses and $\mathcal{L}$ is the set of lines. Also, $\mathcal{N}_{PMU}$ and $\overline{\mathcal{N}}_{PMU}$ are the subsets of buses with and without PMUs respectively, such that $\overline{\mathcal{N}}_{PMU}\cup \mathcal{N}_{PMU}= \mathcal{N}$ and $\overline{\mathcal{N}}_{PMU}\cap \mathcal{N}_{PMU}= \emptyset$. There are $n= |\mathcal{N}|$ buses and $\ell=|\mathcal{L}|$ lines in the power system. Also, $\mathcal{A}_i$ represents the adjacent buses to the $i^{th}$ bus. Fig. \ref{4-bus} exhibits a sample 4-bus grid, where bus 1 is the PMU bus and buses 2, 3, and 4 are non-PMU buses (before adding the PMU 2). In this system, $\mathcal{A}_1 = \{2, 3\}$ and bus 3 is a zero injection bus (ZIB).
\begin{figure}
    \centering
    \includegraphics[width=0.25\textwidth]{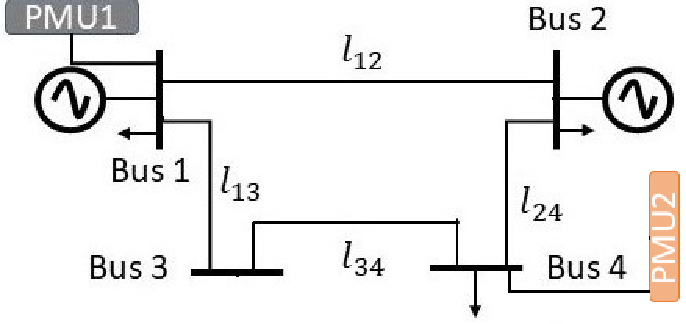}
    \caption{An example power system with four buses and two PMUs}
    \label{4-bus}
    \vspace{-1.5em}
\end{figure}

This work is developed based on DC state estimation, so the states of the system in this work are phase angles denoted by $\theta$. PMUs measure the state of their host buses directly, while the states of the buses in $\overline{\mathcal{N}}_{PMU}$ can be calculated as follows:
\begin{equation}
    \theta_j = \theta_i - p_{ij} x_{ij},
    \label{One}
\end{equation}
where $\theta_i$ is the phase angle of the bus with PMU and $\theta_j,\ j \in \mathcal{A}_i$, is the phase angle of the adjacent bus to the $i^{th}$ bus. Also, $p_{ij}$ represents the flowing power on the line between buses $i$ and $j$ while $x_{ij}$ is the reactance of the line. {Note} that all $p_{ij}$ values are measured values by PMUs which could {be under attack or not}. To make it possible to measure the states of all buses directly or using \eqref{One}, each bus is required to either have a PMU in it or one of its adjacent buses. This requirement is the basis of optimal PMU placement.

\subsection{Optimal PMU placement}
A bus within a power system is considered "observable" under two conditions: either it contains a PMU, or a PMU is installed in any of its adjacent buses. A power system achieves "full observability" when every bus within the system meets the above criteria for observability. Minimizing the following criterion guarantees the full observability of the system while placing a minimum number of PMUs \cite{ghosh2017optimal}: 
\begin{equation}
\min \sum_{m=1}^{n} w_m y_m \hspace{0.5cm} s.t.\hspace{0.3cm} \hspace{0.10cm} g(\textbf{Y})\geq \textbf{B},
\label{Two}
\end{equation}
where $w_m$ is the cost of installing PMU in the $m^{th}$ bus, $\textbf{Y} = [y_1 \hspace{0.2cm} y_2 \hdots y_n]^{T}$ is the binary decision variable for PMU placement with entries:
\begin{equation}
    y_m = 
    \begin{cases}
    1, &  \text{if there is a PMU in the $m^{th}$ bus}\\ 
    0, & \text{otherwise},
    \end{cases}
\end{equation}
also $\textbf{B} = [1\ 1 \hdots 1]^{T}$ of dimension $n$. At last, $g(\textbf{Y})$ is a vector of the same dimension as $\textbf{B}$ and $\textbf{Y}$ of entries:
\begin{equation}
    g_m(\textbf{Y}) = 
    \begin{cases}
    1, & \text{if $m\in\mathcal{N}_{PMU}$ or there is a PMU bus in $\mathcal{A}_{m}$} \\
    0, & \text{otherwise}.
    \end{cases}
    \label{GY}
\end{equation}
Equation \eqref{GY} is a binary function defining whether or not the bus $m$ is observable by at least one PMU.

Power systems can incorporate zero injection buses (ZIBs), which do not contain any load or generation components. According to Kirchhoff's Current Law (KCL), if the power flow is known in all lines connected to a ZIB except for one, then the current in that unknown line can be revealed. Consequently, if all buses connected to a ZIB are observable, applying KCL makes the ZIB observable as well. Furthermore, in the case where all buses connected to an observable ZIB are observable except for one, KCL can also be employed to make the previously unobservable bus observable as well \cite{chen2020pmu}. Optimal PMU placement of power systems with ZIB follows the same steps with just introducing $g_m'(\textbf{Y})$ instead of $g_m(\textbf{Y})$. Algorithm \ref{ZIB-obs} represent the computation of $g_m'(\textbf{Y})$.

\begin{algorithm}
    %\SetKwInOut{Input}{Input}
    %\SetKwInOut{Output}{Output}
    \KwData{$ZIB$, $g_m(\textbf{Y})$}
    \KwResult{$g'_m(\textbf{Y})$}
    \eIf{$g_m(\textbf{Y}) = 1$}
      {
        $g'_m(\textbf{Y}) = 1$
      }
      {
        \eIf{ $\left(g_{k}(\textbf{Y}) = 1, 
         \ \forall k \in \mathcal{A}_m \right) \ \textbf{and} \mbox{} \linebreak
         \left( \mathcal{A}_m\cup \{m\} \right)\ \cap \ ZIB \neq \emptyset $\\}
        {
        $g'_m(\textbf{Y}) = 1$
        }
        {
        $g'_m(\textbf{Y}) = 0$
        }
      }
\caption{Observability of system with ZIB}
\label{ZIB-obs}
\end{algorithm}

In Algorithm \ref{ZIB-obs}, $ZIB$ represents the set of ZIB buses. If an attacker endeavors to alter the value of $\theta_i$ for a PMU bus, as per \eqref{One}, all $\theta_j$ for every $j$ within the set $\mathcal{A}_i$ will also be modified. Consequently, if any bus $j \in \mathcal{A}_i$ is linked to another PMU within its vicinity, any disparity in the calculated or observed phase angles between different PMUs for the same bus will immediately raise an anomaly flag at the defender's end. In contrast, if the attacker manipulates a $p_{ij}$ value, only one phase angle, as computed by \eqref{One}, will be affected. Therefore, this type of attack can go undetected if the bus at the opposite end of the target line also lacks a neighbouring PMU for monitoring purposes.

%While $\theta_i$ and $p_{i,j}$ are true values, $\theta^{meas}_i$and $p^{meas}_{i,j}$ are captured values via measurement tools which could be accurate or not and the adversarial can manipulate these measured values. If the attacker attempts to change the value of $\theta_i^{meas}$ in one of the PMU buses, based on \eqref{One}, all $\theta_j^{meas}, \ \forall j\in \mathcal{A}_i$ will change. As a result, if any bus $j \in\mathcal{A}_i$ has another PMU in its adjacency, the anomaly will be detected due to different phase angles calculated/observed by different PMUs for the same bus. If, on the contrary, the attacker manipulates a $p_{ij}^{meas}$ one phase angle calculated by \eqref{One} changes. Hence, the latter attack will be detected only if the bus on the other end of the target line has another PMU in its adjacency.\\
%If $Pr_j$ is the probability of $j^{th}$ bus to have a second PMU in its adjacency and $j \in \mathcal{A}_i$, then:
%\begin{equation}
 %   Pr_j \leq \sum_{k\in \mathcal{A}_i} Pr_k
%\end{equation}
%\textcolor{magenta}{[V: please fix the paragraph below. needs rewriting.]\\}
In conclusion, although certain attack scenarios have more severe consequences, they also tend to have a higher likelihood of detection.

\subsection{Game theoretic formulation}
\label{sec:game}
In this paper, a game-theoretic approach has been devised to identify the best actions for both attacker and defender and compute the FDIA detection rate based on them.

The strategic interaction between the attacker and defender can be formulated as a two-player zero-sum game \cite{lakshminarayana2021moving}. This game, denoted by $\mathcal{G} = \left(\mathcal{T} \triangleq \{D,A\};\ (\mathcal{S}_D, \mathcal{S}_A);\ (F_D,F_A) \right)$, is composed of the set of players, the sets of possible actions of the players and the utility (or reward or payoff) functions of the players, respectively, and which will be defined next. 

The set of the players of this game is $\mathcal{T} = \{D, A\}$, where $D$ denotes the defender and $A$ the attacker. The set of action profiles for the game is $\mathcal{S} = \mathcal{S}_D \times \mathcal{S}_A$, where $\mathcal{S}_D$ is the set of discrete defense actions and $\mathcal{S}_A$ is the set of discrete attacks. 

The defense action $d \in \mathcal{S}_D \subseteq \overline{\mathcal{N}}_{PMU}$ is the index of nominal buses for a potential additional PMU and it is a subset of buses that do not have a PMU already. Selecting the candidates for installing additional PMU and forming $\mathcal{S}_D$ is based on the knowledge of the defender of the topology of the system.
A set of defensive actions could encompass all non-PMU buses. However, to streamline the process and minimize computational costs, we can exclude buses that cause the same redundant observability as others, thereby reducing the number of actions required. The process of picking candidate buses is detailed in Section \ref{defense_set}. 

The attacker selects a target PMU to manipulate either part or all of its measurements consisting of: the phase angle of the bus and the power flows of connected lines to it. Thus, a specific attack action can be written as $a = (u, \mathcal{V}_{u}),$ in which $u \in \mathcal{N}_{PMU}$ is the index of the bus containing the PMU under attack and $\mathcal{V}_u \in \Pi({\mathcal{P}_u}\cup \{\theta_u\})$ denotes the subset of the measurements of the target PMU at bus $u$ that are manipulated. Also, $\Pi({\mathcal{P}_u}\cup\{\theta_u\})$ represents the set of partitions of ${\mathcal{P}_u}\cup\{\theta_u\}$ containing all measurements of the target PMU and $\mathcal{P}_u =\{ p_{uk}, \forall\ k \in \mathcal{N}\ | \ (u,k)\in \mathcal{L} \} $ is the set of all the power flows $p_{uk}$ of the lines that are connected to bus $u$. The set of all possible attacks can be defined as follows: $\mathcal{S}_A= \{(u,\mathcal{V}_u) \in \mathcal{N}_{PMU} \times \Pi({\mathcal{P}_u}\cup\{\theta_u\})\}$.  The number of such attacks is $|\mathcal{S}_A| = \sum_{u\in \mathcal{N}_{PMU}} (2^{|\mathcal{L}_u|+1} - 1)$.

For the attacked line flows, there are two buses affected: (i) the attacked PMU bus, and (ii) the bus which is at the other end of the line (the line with manipulated power flow value). Also, if a phase angle is manipulated, all of the adjacent buses to the attacked PMU are affected because calculating their phase angles is dependent on the phase angle of the PMU bus. For a precise attack {$a\in \mathcal{S}_A$}, we form the set of affected buses and call it $\mathcal{C}_a$. For example in the 4-bus system depicted in Fig. \ref{4-bus}, the measurements of PMU $1$ that can be tampered with are: $\{ p_{12},p_{13}, \theta_{1} \}$. So the set $\mathcal{S}_A$ for it is thus:
\begin{multline*}
\mathcal{S}_A=\left\{ (1, \{p_{12}\});\ (1, \{p_{13}\});\ (1, \{\theta_1\});\ (1, \{p_{12}, p_{13}\});\right. \\ \left. (1, \{p_{12}, \theta_1\});\ (1,\{p_{13},\theta_1\});\ (1, \{p_{12}, p_{13}, \theta_1\})\right\}.
\end{multline*}
%So, the total number of possible actions for the attacker is $2^3-1=7$.

%\textcolor{magenta}{[V: please replace the $l$ notation with $\ell$ everywhere as above! Actually forget it, since in the above not the lines are modified but the powers of the lines! I've modified accordingly.]}
% \textcolor{red}{[V: please correct the editing bugs! you can't put small caps in the beginning of a sentence and you can't put large caps inside of a sentence! Please print and read and spell check by yourself. Also "$a$s" is NOT proper writing. Since when is a PMU a player?!]}
 %\textcolor{blue}{The defense action $d \in \mathcal{S}_D = \overline{\mathcal{N}}_{PMU}$ is the index of a bus that does not have a PMU and on which the defender places the extra PMU. The number of defenses is hence just $| \overline{\mathcal{N}}_{PMU}|$, the number of buses without PMUs.}
 
The attacker aims to maximize the deviations of phase angles measured by PMUs from their true values. The effect of the FDIA on the phase angles is:
\begin{equation}
    \textbf{E}(a, d) = \Theta -\Theta_{bad},
\end{equation}
%\textcolor{magenta}{[V: I have a question about equation (7): why are not the tampered powers of the lines considered here? Why only the phases? I don't get this! All tempered measurements harm the power system, right?]} 
where $\Theta = [\theta_1 \hspace{0.3cm} \theta_2 \hspace{0.3cm} \hdots \theta_n]^T$ is the vector of accurate phase angles and $\Theta_{bad} = [\theta_{bad,1} \hspace{0.3cm} \theta_{bad,2} \hspace{0.3cm} \hdots \theta_{bad,n}]^T$ is the measured phase angles vector of under attack power system. The defender's goal is to add one extra PMU for the redundant measurement of attack-prone properties in order to detect the possible FDIA.
Let $O_k(d)$ represent the number of PMUs observing the bus $k$, 
\begin{equation}
    O_k(d) =  \left| \mathcal{N}_{PMU} \cap \mathcal{A}_k \right|.
\end{equation}

    {In other words, $O_k(d)$ is the total number of PMUs in the $k^{th}$ bus and its adjacency after the defensive action $d$.} Consider Fig. \ref{4-bus} as an example. When only PMU $1$ {(as per the initial optimal PMU placement)} is deployed within the system, the total number of PMUs observing bus $2$ is equal to $1$. Now, if the operator introduces another PMU in bus $4$ (i.e., $d=4$), {then $O_{2}(4) = 2$ as there will be two PMUs observing bus $2$. In this scenario, a potential anomaly is highlighted in the measurements of at least one of the PMUs if the phase angle calculations from both observing PMUs differ.}

We consider that the attacker's objective is to pick an action $a$ that remains undetected and that maximizes the FDIA effect. To rigorously define the attacker's reward, we introduce $O(a,d)$ as the number of PMUs observing the bus with affected phase angle calculation after the attack and defense actions:
\begin{equation}
    O(a,d) = \left| \mathcal{N}_{PMU} \cap \left\{\bigcup_{i \in \mathcal{C}_a}\mathcal{A}_i\right\} \right|,
\end{equation}
in which $\mathcal{C}_a$ is the set of buses whose phase angle measurements are affected by the attack. If $O(a,d)>1$, then the attack is detected and the reward of the attacker is set to zero (worst case for the attacker). Otherwise, if $O(a,d)=1$, the attack is undetected and its effect is maximized by maximizing {$\|\textbf{E}(a, d)\|$}. 
To sum up, the reward of the attacker is:
\begin{equation}
%\min_{d\in \mathcal{S}_D}\max_{a\in \mathcal{S}_A}\left\{
F_A(a, d) = 
\begin{cases}
\|\textbf{E}(a, d)\|,\hspace{0.2cm} \text{if} \hspace{0.3cm} O(a,d) = 1 \\
0,\hspace{1.5cm} \text{if} \hspace{0.3cm} O(a,d) > 1,
\end{cases}
\label{Obj}
\end{equation}
{which translates that, if only one PMU observes the target bus, the attacker is undetected and can stealthily manipulate measurements, resulting in a reward of $\|\textbf{E}(a, d)\|$. However, if multiple PMUs monitor the target bus, the attack is detected, rendering null the attacker reward.} 

We further consider that the defender's objective is the exact opposite by maximizing
\begin{equation}
F_D(a, d) = - F_A(a, d),
\label{Obj_d}
\end{equation}
which is always negative and maximized when the FDIA attack is detected and $F_D(a,d)=F_A(a,d)=0$.

%Indeed, if $O(a,d)=1$, then the attack is considered successful because the bad data injected cannot be detected. In this case, the attacker wants to maximize the harm it causes to the system by maximizing $\|E(a, d)\|$. The defender has a negative utility in this case as the defense (which is detecting the FDIA) has failed. 
%On the contrary, if $O(a,d)>1$, then the attack is considered unsuccessful because the bad data can be detected. In this case, the attack has failed and the utility of the attacker is zero. 

\subsection{Defensive actions} \label{def act}
\label{defense_set}
%\textcolor{magenta}{[V: where is the selection of the attacker action set? At least a discussion needs to be made here.] }

The defender has full knowledge of the topology of the system and, hence, can use this information to specify more accurately the candidate defense actions by selecting only a subset of $\overline{\mathcal{N}}_{PMU}$. The reason is that the defender wants to add one extra PMU to increase the system's observability in the least observable buses, whereas some buses in $\overline{\mathcal{N}}_{PMU}$ have guaranteed double observability by the PMUs in place. 

Let $\mathcal{B}_\alpha \subseteq \overline{\mathcal{N}}_{PMU}, \hspace{0.2cm} \forall \ \alpha\in\overline{\mathcal{N}}_{PMU}$ denote the subset of non-PMU buses with single observability, which turn double observable after adding a PMU to $\alpha^{th}$ bus. Algorithm \ref{def_action} describes the process of selecting the candidate buses.
\begin{algorithm}
    \SetKwInOut{Input}{Input}
    \SetKwInOut{Output}{Output}
    \Input{$\overline{\mathcal{N}}_{PMU}$, $\mathcal{A}_i$}
    \Output{$\mathcal{S}_D$}
    
    Specify $\mathcal{B}_\alpha$ for all of the non-PMU buses.\\
    Classify the non-PMU buses $\alpha\in\overline{\mathcal{N}}_{PMU}$ as follows: buses $\alpha_1$ and $\alpha_2$ belong to the same class if: either $\mathcal{B}_{\alpha_1} \equiv \mathcal{B}_{\alpha_2}$ or $\mathcal{B}_{\alpha_1} \subset \mathcal{B}_{\alpha_2}$.\\
    From each class, select the one bus with the largest set $\mathcal{B}_\alpha$.\\
    The above-selected buses form $\mathcal{S}_D$.
    \caption{Defining defender's set of actions}
    \label{def_action}
\end{algorithm}
%
%\begin{itemize}
 %   \item \textbf{Step one:} Specify $\mathcal{B}_\alpha$ for all of the non-PMU buses $\alpha\in\overline{\mathcal{N}}_{PMU}$.
  %  \item\textbf{Step two:} Classify the non-PMU buses $\alpha\in\overline{\mathcal{N}}_{PMU}$ as follows; buses $\alpha_1$ and $\alpha_2$ belong to the same class if: either $\mathcal{B}_{\alpha_1} \equiv \mathcal{B}_{\alpha_2}$ or $\mathcal{B}_{\alpha_1} \subset \mathcal{B}_{\alpha_2}$.
%    \item \textbf{Step three:} From each class, select the buses with the largest set $\mathcal{B}_\alpha$.
 %   \item\textbf{Step four:} The above selected buses are the candidate defense actions inside of $\mathcal{S}_D$.
%\end{itemize}

%%%%%%%%%%%%%%%%
%Considering the system presented in figure \ref{4-bus}, while only PMU 1 is in the system, where $\overline{\mathcal{N}}_{PMU} = \{2, 3, 4\}$, $\mathcal{B}_2 = \{2, 4\}$, $\mathcal{B}_3 = \{3, 4\}$, and $\mathcal{B}_4 = \{2, 3, 4\}$. In this example, $\mathcal{B}_{2} \subset \mathcal{B}_{4}$, $\mathcal{B}_{3} \subset \mathcal{B}_{4}$. As a result, all three of these buses form one class and since $\mathcal{B}_{4}$ is the largest set, then bus 4 should be selected as the potential defender action.
%%%%%%%%%%%%%%%%
\section{Robust minimax defense strategy} \label{Solution}

Having defined all the components of the two-player non-cooperative game $\mathcal{G}$ under study, we will now proceed to find the mixed Nash equilibrium solution of this game, which will lead to the robust minimax defense solution in terms of placing one additional PMU against a strategic attacker.

%\subsection{Nash Equilibrium (NE) solution}
\theoremstyle{definition}
\newtheorem{definition}{Definition}
%\textcolor{magenta}{[V: There are several missing pieces here. First, explain what the Nash is intuitively and include the mathematical definition of the Nash equilibrium in mixed strategies. For this you need to introduce the mixed strategies before defining the NE. Second, say that every finite discrete game, hence our game $\mathcal{G}$ as well, has at least one Nash equilibrium in mixed actions and cite J. Nash papers (see our work with Subhash). Third, start discussing about the algorithms to find the NE: i) if full game knowledge is available, Lemke-Howson; in our case this means what exactly? ii) if the players only know their own actions and can measure the payoff value at the end of a game interaction, then we can use EXP3. Please explicit EXP3 acronym when first introduced. Same goes for all acronyms. ]}

The NE is the natural outcome of a non-cooperative game and is a state, or an action profile, from which the players cannot unilaterally deviate without losing in terms of their individual rewards. The mathematical definition of the NE in pure strategies for the game $\mathcal{G}$ under study is given as follows. 

%\textcolor{green!50!black}{A pure strategy Nash equilibrium for the game under study $\mathcal{G}$ is an action profile $(a^*, d^*)$ if with the property that the player $A$ can not do better by choosing an action different from $a^*$ if player $D$ adheres to $d^*$ and player $D$ can not do better by choosing an action different from $d^*$ if player $A$ adheres to $a^*$. In mathematical words}

\begin{definition}\emph{
An action profile $(a^*, d^*) \in \mathcal{S}$ is a NE in pure strategy of the non-cooperative game $\mathcal{G}$, iff  $F_A(a^*, d^*) \geq F_A(a,d^*), \ \forall \ a \ \in \mathcal{S}_A$ and $F_D(a^*, d^*) \geq F_D(a^*,d)$,(or equivalently, $F_A(a^*, d^*) \leq F_A(a^*,d))\ \forall \ d  \in \mathcal{S}_D$.}
\end{definition}

Our finite and discrete game $\mathcal{G}$ might not have a such pure NE solution. Instead, the game {always} has at least one mixed strategy NE solution \cite{fudenberg1991game}. A mixed strategy NE is the solution of the extension of the game to mixed strategies, in which the players choose random actions following certain probability distributions. Therefore, our objective is to find a mixed-strategy NE solution. In our case, the attacker choose a random action $a\in \mathcal{A}$ following a discrete probability distribution $\sigma_A = (\rho_{1}, \rho_{2}, \hdots ,\rho_{|\mathcal{S}_A|}) \in \Delta_A$, such that $\rho_k$ denotes the probability of selecting the $k$-th action in $\mathcal{S}_A$. Similarly, the defender choose a random action $d\in \mathcal{S}_D$ following the discrete probability $\sigma_D = (\mu_{1}, \mu_{2}, \hdots ,\mu_{|\mathcal{S}_D|}) \in \Delta_D$ such that $\mu_k$ denotes the probability of selecting the $k$-th pure defense action in $\mathcal{S}_D$. We will also make use of the notations $\rho_a$ and $\mu_d$ to denote the probabilities of selecting arbitrary actions $a \in \mathcal{S}_A$ and $d\in \mathcal{S}_D$, respectively. At last, the sets $\Delta_A$ and $\Delta_D$ are the corresponding discrete probability distribution simplices:
$$\Delta_A = \left\{\sigma_A=(\rho_{1}, \hdots ,\rho_{|\mathcal{S}_A|}) \in [0,1]^{|\mathcal{S}_A|} \ \left| \ \sum_{k=1}^{|\mathcal{S}_A|}  \rho_{k} = 1 \right. \right\}$$
$$\Delta_D = \left\{\sigma_D=(\mu_{1}, \hdots ,\mu_{|\mathcal{S}_D|}) \in [0,1]^{|\mathcal{S}_D|} \ \left| \ \sum_{k=1}^{|\mathcal{S}_D|}  \mu_{k} = 1 \right. \right\}.$$

 The modified rewards of the extended game are the mathematical expectations of the obtained rewards given the randomly chosen actions that follow the distribution $\sigma_A$ for the attacker and $\sigma_D$ for the defender:
\begin{equation}
    \hat{F}_A(\sigma_A, \sigma_D) = \sum_{a \in \mathcal{S}_A}\sum_{d \in \mathcal{S}_D} F_A(a, d) \ \rho_{a} \ \mu_{d}.
\end{equation}

%\begin{definition}
%  A mixed strategy profile $(\sigma_a^*, \sigma_d^*)$ is a mixed NE if of the game $\mathcal{G}$, iff it is a NE of the extended game, i.e., $\hat{F}_A(\sigma_a^*, \sigma_d^*)\geq \hat{F}_A(a, \sigma_d^*), \ \forall a \in \mathcal{S}_A$ and $ \hat{F}_D(\sigma_a^*, \sigma_d^*)\geq \hat{F}_D(\sigma_a^*, d), \ \forall d \in \mathcal{S}_D$.
%\end{definition}

To sum up, the extended game to mixed strategies can be defined as $\hat{\mathcal{G}} = \left(\mathcal{T} \triangleq \{D,A\}; (\Delta_D, \Delta_A); (\hat{F}_D,\hat{F}_A) \right)$ {and the mixed strategy NE is defined as follows.}

\begin{definition}\emph{
  A mixed strategy profile $(\sigma_a^*, \sigma_d^*)\in \Delta_A \times \Delta_D$ is a mixed NE of the game $\mathcal{G}$, iff it is a NE of the extended game $\hat{\mathcal{G}}$ such that $\hat{F}_A(\sigma_A^*, \sigma_D^*)\geq \hat{F}_A(\sigma_A, \sigma_D^*), \ \forall \ \sigma_A \in \Delta_A$, and $ \hat{F}_D(\sigma_A^*, \sigma_D^*)\geq \hat{F}_D(\sigma_A^*, \sigma_D), \ \forall \ \sigma_D \in \Delta_D$.}
\end{definition}

The mixed strategy NE can be calculated via the Von-Neumann indifference principle \cite{fudenberg1991game} by solving a certain number of linear systems of equations and inequalities. {This number} grows exponentially with the number of actions of the players. When the number of actions grows large, finding the NE via the Von-Neumann indifference principle becomes intractable. The Lemke-Howson method \cite{roughgarden2010algorithmic} is the most efficient alternative known to date. However, in the worst cases, the complexity of the Lemke-Howson algorithm is the same as the Von-Neumann indifference principle.

Both the Von-Neumann indifference principle and the Lemke-Howson method require full knowledge of the payoffs and sets of actions of players. Lack of knowledge about the payoffs of the opponent player hinders the game from being solved by them. Inspired by \cite{lakshminarayana2021moving}, the NE is found using the EXP3 algorithm from the multi-armed bandits framework.
The main desirable feature of EXP3 is that neither player is required to have full knowledge of the game.

Indeed, in EXP3 algorithm, the agent $A$ or $D$ draws a random action $a_t$ or $d_t$ at each iteration $t$ from the distribution $\sigma_{A,t}=(\rho_{1,t}, \rho_{2,t}, \hdots ,\rho_{|\mathcal{S}_A|,t})$ or $\sigma_{D,t}=(\mu_{1,t}, \mu_{2,t}, \hdots ,\mu_{|\mathcal{S}_D|,t})$ and observes its own resulting reward. The rewards of the non-chosen actions are not known and have to be estimated. For instance, at the attacker side, the estimated rewards are as follows: \footnote{Similar equations can be written for the defender and are omitted here.}
\begin{equation}
    \hat{F}_{A, t} (a) = \frac{F_A (a_t, d_t)\mathds{1} [a = a_t] + \beta_t}{\sigma_{A, t}(a)}, \ \forall a \in \mathcal{S}_a
\end{equation}
where $\beta_t > 0$ controls the estimator's variance. Additionally, {$\sigma_{A, t}(a)$} signifies the probability {of choosing action $a$ at iteration $t$}. The next equation captures the cumulative score of these actions and determines the performance of actions in the past.
\begin{equation}
    G_{A,t} (a) = \sum_{\tau = 1}^t \eta_\tau \hat{F}_{A, \tau} (a)
\end{equation}
where $\eta_\tau > 0$ is a learning parameter. Then, {the updated probability distribution $\sigma_{A, t+1}$ is computed,}
\begin{equation}
    \sigma_ {A, t+1} (a) = \gamma_t \frac{1}{|\mathcal{S}_A|} + (1 - \gamma_t)\frac{\exp(G_{A,t} (a))}{\sum_{r \in \mathcal{S}_A} \exp(G_{A,t} (r)}, \forall a,
\end{equation}
where $\gamma_t \in (0, 1]$ is another learning parameter. Similarly, the defender updates its own probability distribution. 
The process repeats until convergence. According to \cite{hart2000simple}, the empirical frequency of actions of EXP3, defined below, converges to the NE.
\begin{equation}
    \Bar{\sigma}_{A, t} = \frac{1}{\sum_{\tau = 1}^t \eta_\tau}\sum_{\tau = 1}^t \eta_\tau \sigma_{A, \tau}.
\end{equation}

\section{Numerical results} \label{result}

In this section, we evaluate the proposed framework for the IEEE 14-bus system. The initial optimal PMU locations for satisfying the full observability condition are chosen as in \cite{ghosh2017optimal}. {In this grid, as the result of optimal PMU placement, buses 2, 6, 7, and 9 while not considering ZIB and buses 2, 6, and 9 with considering ZIB are PMU buses. Note that in this test system, $ZIB=\{7\}$ as bus $7$ is the ZIB.} Also, Following the Section \ref{def act}, the set of candidate buses to place the additional PMU is $\mathcal{S}_D = \{1, 3, 8, 10, 13\}$. 
%Table \ref{Locs} shows the obtained optimal PMU locations with and without considering ZIB. 

Additionally, the NE has been calculated by the Lemke-Hawson method with full game knowledge and compared to the EXP3 algorithm, which is expected to converge to the NE.

% \begin{figure}
%     \centering
%     \includegraphics[width=0.3\textwidth]{14-bus.jpg}
%     \label{14bus}
%     \caption{IEEE 14-bus system}
%     \label{14bus}
%     \vspace{-1em}
% \end{figure}
 % \begin{table}[h]
 % \vspace{-0.2in}
 %     \caption{Optimal Locations of PMUs in IEEE 14-bus system}
 %     \centering
 %     \begin{tabular}{|c|c|} \hline       
 %         \multicolumn{2}{|c|}{ Optimal Buses for PMUs} \\
 %        \hline \hline
 %        Without ZIB & With ZIB \\
 %        \hline
 %         2, 6, 7, 9 & 2, 6, 9\\
 %         \hline
 %     \end{tabular}
 %     \label{Locs}
 %     \vspace{-1.5em}
 % \end{table}
%\subsection{Specification of the defender's set of actions}
\subsection{Evaluation of the NE solution}
We first compute the NE solution via Lemke-Howson and EXP3 algorithms for comparison purposes.
Tables \ref{lemke1} and \ref{LM-ZIB} contain the {probability distributions of NE for defender and attacker}. Additionally, they contain the results calculated in the last iteration of the EXP3. The provided values highlight the fact that EXP3 enables us to compute the mixed strategy NE without requiring full knowledge about the set of actions and payoffs {of the opponent}. indeed, for two-player zero-sum games, the EXP3 algorithm converges to the NE solution \cite{lazaric}.
Now, to assess the performance of the NE solution in terms of FDIA detection rate, we introduce the probability of detecting and not detecting an attack given a randomly chosen defense action $d\in \mathcal{S}_D$ and attack action $a \in  \mathcal{S}_A $. Since the actions of the attacker and the defender are independent of one another, we can calculate the probability of detecting and not detecting FDIA as follows:
\begin{equation}
\mathrm{Pr}[O(a,d)>1]= \sum_{a\in \mathcal{S}_A,\ d\in \mathcal{S}_D} \mathrm{Pr}(a)\ \mathrm{Pr}(d) \ \mathds{1}_{[O(a,d)>1]}, 
\end{equation}
and $\mathrm{Pr}[O(a,d)=1] = 1- \mathrm{Pr}[O(a,d)>1]$ given that $O(a,d)\geq 1$ always, where {$\mathds{1}_{[t]}$} is the indicator function that equals one if the condition $t$ is true and zero otherwise. 

At the NE $(\sigma_A^*, \sigma_D^*)$, the attack detection rate is given by:
\begin{equation}
\mathrm{Pr}[O(a,d)>1]= \sum_{a\in \mathcal{S}_A,\ d\in \mathcal{S}_D} \rho_a^*\  \mu_d^* \ \mathds{1}_{[O(a,d)>1]}.
\end{equation}
The above will be compared with a naive defense strategy where the probability of choosing an action $d\in \mathcal{S}_D$ is uniformly distributed: $\mathrm{Pr}(d) = 1/|\mathcal{S}_D|, \forall \ d \in \mathcal{S}_D$. For this naive defense strategy, the attack detection rate is given by:
\begin{equation}
\mathrm{Pr}[O(a,d)>1]= \frac{1}{|\mathcal{S}_D|}\sum_{a\in \mathcal{S}_A,\ d\in \mathcal{S}_D} \rho_a^*\   \ \mathds{1}_{[O(a,d)>1]},
\end{equation}
\newgeometry{bottom=0.7in} % Adjust bottom margin for this page
% Place your equation or content here
\restoregeometry % Restore original margins
assuming the attacker follows its NE via the $\sigma_A^*$ probability distribution. 

The assessment of FDIA detection rate should involve evaluating their performance against intelligent attacks. Note that none of the conducted FDIAs can be detected without the implementation of a defensive measure, such as the addition of an extra PMU, as outlined in this paper.

\begin{table}
    \centering
        \caption{NE solution for the IEEE 14-bus system without ZIB calculated via the Lemke-Howson and EXP3 methods}
    \begin{tabular}{ccc} \toprule
  \multirow{2}{4em}{Action (Attacker)\centering} & \multicolumn{2}{c} {NE probability distributions} \\ \cline{2-3}
         & Lemke-Howson & EXP3 \\ \midrule
        Line 1-2 & 0.3113 & 0.3220 \\ 
        Line 2-3  &0.4923 &0.4726 \\
        Lines 6-11, 6-12, 6-13  &0.1965 &0.2044 \\
        Other Actions & 0 & $\approx0$\\ \bottomrule
        Action (Defender) & &\\ \midrule
        Bus 1 & 0.3774 &0.3758 \\ 
        Bus 3 & 0.6071&0.6062 \\ 
        Bus 10 &0.0155 & $<0.01$\\
        Other Actions & 0 & $ \approx 0$\\  \bottomrule  
    \end{tabular}
    \vspace{-1.5em}
     \label{lemke1}
\end{table}

\begin{table}
    \centering
    \caption{NE solution for IEEE 14-bus system with ZIB calculated via Lemke-Howson and EXP3 methods}
    \begin{tabular}{ccc} \toprule
  \multirow{2}{4em}{Action (Attacker)} & \multicolumn{2}{c} {NE probability distributions} \\ \cline{2-3}
         & Lemke-Howson & EXP3 \\ \midrule
       Line 2-3 & 0.7685 & 0.7401\\ 
        Lines 6-11, 6-12, 6-13 & 0.2315 & 0.2254\\ 
      Other Actions & 0 & $<0.01$ \\ \bottomrule
        Action (Defender) & &\\ \midrule
        Bus 3 & 0.7685 & 0.7769\\ 
        Bus 10 & 0.2032 & 0.1584\\ 
        Bus 13 & 0.0283 & 0.0630\\ 
        Other Actions & 0 & $<0.01$ \\ \bottomrule
    \end{tabular}
    
    \label{LM-ZIB}
    \vspace{-1.5em}
\end{table}

Using the obtained NE solution (in Table  \ref{lemke1} and \ref{LM-ZIB}), the probability of detecting FDIA has been calculated and presented in Table \ref{d-rate} in comparison with the naive defense described above. Adding the PMU based on the proposed method in this paper results in $40.75\%$ and $62.50\%$ detection rates in systems without and with ZIB, respectively. However, adding the PMU with the naive process with the uniformly distributed probability for candidate buses as defender's action results in $25.90\%$ and $23.81\%$ detection rates. So, the benefit of following the proposed algorithm is $14.85\%$ and $36.69\%$ improvements in detection rates with the same number (one) of additional PMU.
\begin{table}[h]
    \centering
        \caption{\small Robust NE defense strategy and a naive defense against a strategic attacker.}
    \begin{tabular}{p{2cm}p{1.6cm}p{1.5cm}p{1.2cm}}   \toprule  
    & Defense type & Without ZIB & With ZIB\\ \midrule
      \multirow{2}{8em}{FDIA detection rate $(\%)$}&\multirow{1}{4em} {Naive\centering} &\multirow{1}{4em} {25.90} & \multirow{1}{4em} {23.81} \\ 
      & \multirow{1}{5em}{Robust NE\centering} &\multirow{1}{5em} {40.75} &\multirow{1}{4em} {62.50} \\ \bottomrule
    \end{tabular}
    \label{d-rate}
    \vspace{-1.5em}
\end{table}
\section{Conclusions} \label{conclusions}
In this paper, after the first optimal PMU placement stage, a two-player zero-sum non-cooperative game is introduced to find a robust defense solution against FDIA by including a single additional PMU. The two players (attacker and defender) have opposite objectives, and neither side has complete information about the game (e.g., the opponent's actions). A reinforcement learning approach called ``EXP3" is exploited to compute the robust Nash equilibrium solution. Our results show that the proposed method increases the rate of FDIA detection while being cost-efficient and robust to strategic attacks which encourages the expansion of this work to larger systems, which may require more than one PMU to reach a certain level of detection.

\bibliography{IEEEabrv,Refs}

\end{document}